\documentclass{PoS}

\title{Universality and Scaling at the chiral transition in 
       two-flavor QCD at finite temperature}

\ShortTitle{Universality and Scaling at the two-flavor-QCD transition}

\author{Tereza Mendes \\
        Instituto de F\'\i sica de S\~ao Carlos, Universidade de S\~ao Paulo, \\
        Caixa Postal 369, 13560-970 S\~ao Carlos, SP, Brazil \\
        E-mail: \email{mendes@ifsc.usp.br}}

\abstract{
The order of the phase transition in finite-temperature QCD with two 
degenerate light quarks is still an open problem and corresponds to
the last question mark in the zero-density phase diagram of QCD.
We argue that establishing the nature of the transition in this case
is also a crucial test for numerical simulations of lattice QCD,
allowing precise estimates of possible systematic errors related
e.g.\ to the choice of fermion-simulation algorithm or of discretized
formulation for fermions.
}

\FullConference{The XXV International Symposium on Lattice Field Theory\\
		 July 30-4 August 2007\\
		 Regensburg, Germany}

\begin{document}

%%%%%%%%%%%%%%%%%%%%%%%%%%%%%%%%%%%%%%%%%%%%%%%%%%%%%%%%%%%%%%%%%%%%%%%%%%%%%%%%%%%%%%%%

\section{Introduction}

The phase diagram at zero baryon density and varying quark masses has been
intensively investigated by lattice simulations \cite{Heller:2006ub}. 
The case of two dynamic quarks, i.e.\ considering dynamic effects
of only two degenerate light-quark flavors, corresponding to the up and down 
quarks, is particularly interesting. 
In this case, if the transition is of second order, 
one would expect to observe universal critical
scaling in the class of the $3d$ $O(4)$ continuous-spin 
model \cite{pisarsky}.
Also, in the continuum limit, simulations using different discretizations
for the fermion fields should give the same results.
The fact that the critical behavior should be in the universality class of a
spin model can be precisely checked, since the nonperturbative behavior for
these models can be obtained with Monte Carlo simulations by so-called {\em global}
methods, which avoid the critical slowing-down present in QCD 
simulations.

The determination of the correct nature of the transition in the two-flavor
case is one of the present challenges of lattice QCD.
In fact, it corresponds to the last question 
mark in the zero-density phase diagram (see e.g.\ \cite{Heller:2006ub}).
This prediction has been investigated numerically by lattice simulations
for almost twenty years, yet there is still no agreement about the order of
the transition or about its scaling properties.
More precisely, the predicted $O(4)$ scaling has been observed
in the Wilson-fermion case \cite{wilson},
but not in the staggered-fermion
case, believed to be the appropriate formulation for studies of the
chiral region. In this case, extensive numerical studies and scaling tests
have been done in the past by the Bielefeld \cite{karsch}, JLQCD \cite{aoki} and 
MILC \cite{bernard} groups. It was found that the chiral-susceptibility peaks
scale reasonably well with the predicted exponents, but no agreement is
seen in a comparison with the $O(4)$ scaling function.
At the same time, some recent numerical studies with staggered fermions
claim that the deconfining transition may be of first order \cite{delia,Cossu:2007mn}.
On the other hand, studies of the two-flavor case in the massless limit point towards
a second-order transition in the $O(2)$ universality class \cite{Kogut:2006gt}.
In summary, the issue of the nature of the QCD transition for the two-flavor
zero-density case is still an open problem, just as it was ten years 
ago \cite{Laermann:1998gf}.

Although scaling is not observed, a simple method may be 
used \cite{conf6,Mendes:2006zf}
in order to obtain a uniquely defined normalization of the QCD data,
allowing an unambiguous comparison to the (normalized) $O(4)$ scaling
function.
This analysis shows a surprisingly better agreement for the
{\em larger} values of the quark masses. Let us note that in previous 
scaling tests the comparison had been done up to a (non-universal) 
normalization of the data and a match to the scaling function was 
tried by fitting it to the data points with the smallest masses.
One interpretation of this result
is that data at smaller masses (closer to the physical
values) suffer more strongly from systematic errors in the simulations.
In fact, larger quark masses are much easier to simulate, allowing 
greater control over errors and more reliable results. 
Here we present a study at a rather large value for the 
light quark mass $m_q$ (we take $m_q = 0.075$ in lattice units), 
using staggered fermions and the MILC code.
We consider the standard (i.e., unimproved) action and temporal lattice 
extent $N_{\tau}=4$, as in most of the studies mentioned above.
Also, we use the so-called R algorithm, but consider very small integration 
steps (of about $m_q/10$).

\section{Scaling tests}
\label{univ}
The behavior of systems around a second-order phase transition 
(or critical point) may show striking similarities for systems 
that would otherwise seem completely different. In fact, it is
possible to divide systems into so-called universality classes,
in such a way that each class will have, e.g., the same critical
exponents around the transition. The critical point corresponds to
external magnetic field $H$ equal to zero, and temperature $T$
given by its critical value $T_c$. Typical exponents are
\begin{eqnarray}
M_{h=0,\,t\to 0^-} &\to & |t|^{\beta}
\mbox{,} \\
\chi_{h=0,\,t\to 0} &\to & |t|^{-\gamma}
\mbox{,} \\
M_{t=0,\,h\to 0} &\to & h^{1/\delta}\,,
\end{eqnarray}
where $M$ is the order parameter --- e.g.\ the magnetization
for a spin system --- $\chi$ is the corresponding susceptibility and
$t = (T-T_c)/T_0$, $h = H/H_0$
are the reduced temperature and magnetic field, respectively.
($T_0$ and $H_0$ are normalization constants.)

Thus, in principle, one may compare the critical exponents for
different systems to check if they belong to the
same universality class. In practice, however, the critical exponents
may vary little from one class to the other and in order to carry out
the comparison one would need to have a very precise determination
of the exponents, which is not yet feasible in the QCD case.

A more general comparison is obtained through the {\em scaling functions}
for both systems. This comparison allows a more
conclusive test, and can be applied for cases where the critical 
exponents cannot be established with great accuracy.
In this case we may assume the exponents for a given class and 
compare the behavior of the whole critical region for one system
to the known scaling curve for the proposed universality class.
The scaling Ansatz is written for the free energy $F_s$ in
the critical region as
\begin{equation}
F_s(t,h) \;=\;
b^{-d}\,F_s(b^{y_t}\,t, b^{y_h}\,h)\,,
\label{ansatz}
\end{equation}
where $b$ is a rescaling factor, $d$ is the dimension and
$y_t,y_h$ are related to the usual
critical exponents: $\beta$, $\gamma$, $\delta$ mentioned above.
Correspondingly, the order parameter
must be described by a universal function
\begin{equation}
M/h^{1/\delta} = 
f_M(t/h^{1/\beta \delta})\;.
\end{equation}
The statement that the function $f_M$ is {\em universal} means that
once the non-universal normalization constants $T_0$
and $H_0$ are determined for a given system in the universality class,
the order parameter $M$ scales according to the {\em same}
scaling function $f_M$ for all systems in this class.
As said above, the comparison of (normalized) scaling functions 
between two systems is a more general test of universality, especially 
in the case of the QCD phase transition.

A further difficulty in studying the critical behavior at the QCD phase
transition is the impossibility of considering the critical point directly,
since that would correspond to having zero quark mass, or
zero magnetic field $H$ in the language of the spin models above.
In order to check scaling with critical exponents of a given class,
or to determine the normalization constants $T_0$ and $H_0$
for systems where a study at $H=0$ is not
possible, it is important to determine the
{\em pseudo-critical line}, defined by the points where
the susceptibility $\chi$ shows a (finite) peak. This
corresponds to the rounding of the divergence that would be observed
for $H=0$, $T=T_c$. The susceptibility scales as
\begin{equation}
\chi \,=\, \partial M/\partial H \,=\,
(1/H_0)\,h^{1/\delta - 1} 
\,f_{\chi}(t/h^{1/\beta})\;,
\end{equation}
where
$f_{\chi}$ is a universal function related to $f_M$.
At each fixed $h$ the peak in $\chi$ is given by
\begin{eqnarray}
t_{p} &=& z_p\,h^{1/\beta \delta}, \\
M_p &=& h^{1/\delta}\,f_M(z_p), \\
H_0\,\chi_{p} &=& h^{1/\delta - 1} \, f_{\chi}(z_p)\,.
\end{eqnarray}
Thus, the behavior along the pseudo-critical line is determined by the   
universal constants
$z_p$, $f_M(z_p)$, $f_{\chi}(z_p)$.
Critical exponents, the scaling function $f_M$
and the universal constants above are well-known for the 
$3d$ $O(4)$ model \cite{o4,o4o2,o4new}.

Note that one may also consider the comparison for finite-size-scaling
functions, since they are also universal and have
the advantage of being valid for finite values of $L$, the linear size
of the system. Such functions can be determined numerically \cite{o4o2}
for the $3d$ $O(4)$ model.

\section{Comparison of QCD data with the predicted scaling function}

We now turn to the comparison of the two-flavor QCD data in the critical 
region (in the case of small but nonzero quark mass) to the predicted
scaling properties of the $3d$ $O(4)$ spin model.
As mentioned in the Introduction, we consider the chiral phase transition,
since there is no clear order parameter for the deconfinement transition
in the case of full QCD.
The order parameter for the chiral transition is given by the so-called
chiral condensate $<\overline \psi \,\psi>$, where $\psi$ is a combination
of the quark fields entering the QCD Lagrangian.
The analogue of the magnetic field 
is the quark mass $m_q$, and (on the lattice) the reduced 
temperature is proportional to $\,6/g^2 - 6/g_c^2(0)\,$,
where $g$ is the lattice bare coupling and $g_c^2(0)$ is its extrapolated
critical value.
Therefore, referring to the pseudo-critical line described in the
previous section, the chiral susceptibility peaks at
\begin{equation}
t_{p}\sim {m_q}^{1/\beta\delta}\,.
\end{equation}

As mentioned in the Introduction, previous results from lattice-QCD 
simulations in the two-flavor case show good scaling (with the predicted 
exponents) {\em only} along the pseudo-critical line, which is given by the 
peaks of the chiral susceptibility. It should be clear from the discussion
in the above sections that this is not a sufficient test to prove that
the transition is second order, especially if no agreement is seen
when comparing the data to the scaling function.
We use \cite{Mendes:2006zf} the observed scaling along the pseudo-critical 
line and the universal quantities $z_p$, $f_M(z_p)$ from the $O(4)$ model
to determine the normalization constants
$H_0$, $T_0$ for the QCD data. This allows an unambiguous 
comparison of the data to the scaling function $f_M$. 
More precisely, we note that in previous analyses
the normalization constants were tentatively adjusted
by shifting the $O(4)$ curve so as to get a rough agreement
with the data at smaller quark masses,
since these are closer to the chiral limit. 
The problem is that the lighter masses are also more subject to
the presence of systematic errors in the simulations.
In this case the overall
agreement was rather poor, indicating that there were strong systematic 
effets or that the transition is not in the predicted universality class.
Here we fix the constants as described in Section \ref{univ},
following the behavior along the pseudo-critical line. 
In this way no value of the quark mass is priviledged and the comparison 
is unambiguous. 

Our comparison \cite{Mendes:2006zf} is shown in Fig.\ \ref{scaling} below.
We note that the integration step for the data at the largest 
mass is considerably smaller than the usual values.
(Our runs were done with 7000 trajectories and 125 steps of
length 0.008 per trajectory.)
The pseudo-critical line corresponds to a single point in this plot and
is marked with an arrow. For clarity we do not show the data --- 
from the Bielefeld and JLQCD collaborations --- obtained directly at
the pseudo-critical point. These are slightly scattered around $z_p$ but
show good scaling within errors.

\begin{figure}[htbp]
\includegraphics[width=0.9\textwidth]{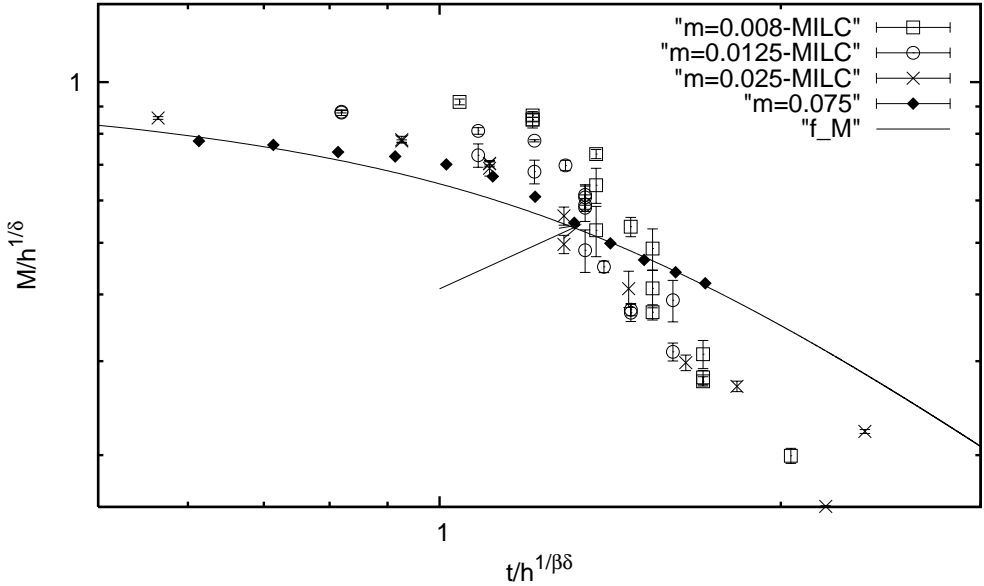}
\caption{
Comparison of QCD (staggered) data to the $O(4)$ scaling
function \protect\cite{Mendes:2006zf}. For clarity, we do not show the data around
the pseudo-critical point (indicated by the arrow), which were
used to determine the normalization of the remaining data 
points.}
\label{scaling}
\end{figure}
We see relatively good scaling in the pseudo-critical region,
i.e.\ around [$z_p$, $\,f_M(z_p)$], as expected.
Away from this region most MILC points
are several standard deviations away from the predicted curve.
These data are given for three values of the quark mass in lattice units:
0.008, 0.0125 and 0.025.
Note that the points with larger mass come closer to the curve.
In particular, we can see that the new data at $m_q = 0.075$ show sensibly
better scaling, especially for larger temperatures, where previously
the scaling seemed unlikely.
The good agreement of these data with the $O(4)$ scaling function
motivates a careful study of systematic errors for smaller masses.
A possible source of such errors are finite-size corrections, which would be
stronger for smaller masses, since then the lattice side may not be large
enough to ``contain'' the physical particle.
Put differently, finite-size effects are expected when the correlation length 
(in lattice units) associated with a particle is comparable
to or larger than the lattice side. Of course, this is more likely to occur
for a lighter particle. 

We note that a finite-size-scaling analysis may also
be carried out using universal functions from the $O(4)$ case
(see \cite{Mendes:2006zf} and references thereof),
but one finds that the QCD data show good (finite-size) 
scaling only along the pseudo-critical line. Moreover, with the
temperature and mass parameters considered, the QCD data 
lie within the asymptotic region, where infinite-volume behavior 
should already be seen. In any case, the finite-size scaling 
away from the pseudo-critical region is also significantly better for
the larger quark masses.

\section{Conclusions}
Determining the nature of the chiral phase transition in
two-flavor QCD still stands as a challenge. Despite great computational
effort, the prediction of a second-order transition with critical behavior
in the universality class of the $O(4)$ spin model is not verified for
staggered fermions of small masses, although it can be shown (by
an unambiguous normalization of the data) that better scaling is 
obtained for the existing data at larger (unphysical) masses.
Let us also mention that a redefinition of the reduced temperature in terms 
of the physical temperature $T$ including \cite{delia} a term in the quark mass 
$m_q$, improves the agreement with the scaling curve further \cite{conf6}.

The fact that data for heavier quarks would show such good scaling
may be surprising, since the normalization of the data for comparison with
the scaling curve did not priviledge any particular values of the quark mass.
Indeed, one would expect to observe agreement with the $O(4)$ scaling curve
for the smaller masses, which are closer to the chiral limit, where universality 
is predicted to hold.
This suggests that the lack of scaling at small masses observed so far may be 
caused by systematic effects. A common source of systematic errors in lattice
simulations are discretization effects, which might be related here to the small
number of points along the temporal direction ($N_{\tau}=4$) or to the use of
the unimproved fermion action. The use of the staggered fermion formulation has 
also been generally criticized recently \cite{Sharpe:2006re}.
Other sources of errors could be finite-size corrections and
uncontrolled errors in the hybrid Monte Carlo algorithm (the R algorithm,
in our case) used for updating the configurations. Both these sources of errors 
would be more significant for the case of smaller masses.

We note that, as mentioned above, the deviations
from $O(4)$ scaling at smaller masses 
are most likely {\em not} due to finite-size corrections.
On the other hand, it must be stressed that the R algorithm used to update 
the configurations 
is not exact and should have its accuracy tested carefully for each different 
value of the quark mass used, comparing when possible to simulations using the
exact RHMC algorithm.
Indeed, recent studies try to establish to what extent the use of the R algorithm 
may have influenced currently accepted results from lattice-QCD simulations.
One study of this type considers possible systematic effects on results for 
the finite-temperature phase transition
in the case of a realistic mass spectrum \cite{Cheng:2006aj}.
(Note that the transition in this case is not of second order.)
It is generally believed that such systematic errors are negligible, or of
the order of the statistical uncertainties in the simulations
\cite{Cossu:2007mn}, but a detailed comparison must be made.
We are currently extending our study using the RHMC algorithm.

As discussed here, the phase transition for two degenerate quark 
flavors is clearly especially well-suited for studying the effects of
systematic errors in lattice simulations, since very large 
systematic effects would not be unexpected around a second-order phase transition.
In fact, this case may be a more stringent test of the several possible systematic
errors, such as discretization effects, errors associated with the updating 
algorithm or with the choice of the fermion discretization used in the simulations.

\section{Acknowledgements}
This work was partially supported by FAPESP and CNPq.

\end{document}